\documentclass[fleqn,usenatbib]{mnras}

\usepackage{newtxtext,newtxmath}

\usepackage[T1]{fontenc}

\DeclareRobustCommand{\VAN}[3]{#2}
\let\VANthebibliography\thebibliography
\def\thebibliography{\DeclareRobustCommand{\VAN}[3]{##3}\VANthebibliography}

\usepackage{graphicx}	
\usepackage{amsmath}	

\title[Life on the Edge]{Life on the Edge: Using Planetary Context to Enhance Biosignatures and Avoid False Positives}
\author[R. Arthur, A. E. Nicholson and N. J. Mayne]{
R. Arthur,$^{1}$\thanks{E-mail: R.Arthur@exeter.ac.uk}
A. E. Nicholson$^{2}$
and  N. J. Mayne$^{2}$
\\
$^{1}$Department of Computer science and $^{2}$Department of Physics and Astronomy, Faculty of Environment Science and Economy, University of Exeter, EX4 4QL, UK.
}

\date{Accepted XXX. Received YYY; in original form ZZZ}

\pubyear{\the\year{}}

\begin{document}
\label{firstpage}
\pagerange{\pageref{firstpage}--\pageref{lastpage}}
\maketitle

\begin{abstract}
We use a probability theory framework to discuss the search for biosignatures. This perspective allows us to analyse the potential for different biosignatures to provide convincing evidence of extraterrestrial life and to formalise frameworks for accumulating evidence. Analysing biosignatures as a function of planetary context motivates the introduction of `peribiosignatures', biosignatures observed where life is unlikely. We argue, based on prior work in Gaia theory, that habitability itself is an example of a peribiosignature. Finally, we discuss the implications of context dependence on observational strategy, suggesting that searching the edges of the habitable zone rather than the middle might be more likely to provide convincing evidence of life.
\end{abstract}




\section{Introduction}\label{sec:introduction}

\begin{quote}\emph{When you have eliminated the impossible, whatever remains, however improbable, must be the truth} - \cite{acdoyle_1885}\end{quote}

The search for alien life is conducted through the search for biosignatures. These are \emph{any substance, group of substances, or phenomenon that provides evidence of life} \citep{catlingdavid2018exoplanet}. A large number of potential biosignatures have been suggested, which are reviewed in e.g. \citet{schwieterman2018exoplanet,chou2024chapter} and compiled in the Life Detection Knowledge Base \citep{ldkb2025}. Many authors \citep[e.g.,][]{spiegel2012bayesian, catlingdavid2018exoplanet, walker2018exoplanet} advocate a Bayesian framework as a way of evaluating biosignatures. In this paper we shift the focus instead towards minimising false positives, cases where biosignatures are detected but no life is present. We also discuss how planetary context, by which we mean planetary features like distance from the host star, size, age and so on, can affect the `quality' of biosignatures and hence the probability of a false positive.

The emphasis on false positives is particularly apt for life detection as it formalises accepted notions of when we can confidently declare that life has been discovered, namely that life is the `hypothesis of last resort' \citep{sagan1993search,catlingdavid2018exoplanet}. This perspective, as also discussed by many others \citep{vickers2023confidence, smith2023life, foote2023false}, derives from a long history of overly optimistic claims of extraterrestrial life detection. Famous examples include Martian canals \citep{lowell1909mars}, Martian vegetation \citep{sinton1957spectroscopic} and biogenic markers in the Martian meteorite ALH84001 \citep{mckay1996search}. Other biosignatures which were once thought to be very reliable, or `smoking guns' for life detection are also now disputed. For example, \citet{krissansen2021oxygen} discussed oxygen accumulation in planetary atmospheres to levels comparable to modern Earth via abiotic pathways alone. One of the foundational atmospheric biosignatures, departures from chemical equilibrium \citep{hitchcock1967life,krissansen2018disequilibrium}, has been shown to occur in abiotic systems \citep{wogan2020chemical}. Recently, phosphine, a gas without a known mechanism of abiotic synthesis on Earth, was proposed as a biosignature \citep{seager2012astrophysical} and was subsequently detected in the atmosphere of Venus \citep{greaves2021phosphine}. However, due to the extreme conditions on Venus, unknown abiotic explanations may remain a more plausible explanation than life, see \citet{bains2021phosphine} for a detailed discussion. Another, even more recent example is the potential detection of DiMethyl Sulfide in the atmosphere of K2-18b which, in turn, has been posited as evidence of life \citep{madhu2025}. However, the signal itself is contested \citep{taylor2025}, abiotic pathways to its production revealed \citep{reed2024,nora2024} and even the overall nature of the planet K2-18b uncertain \citep{wogan2024}. This example clearly illustrates that regardless of whether one is agnostic to the presence of life or otherwise on K2-18b, ruling out false positives is, perhaps, the most challenging aspect. 

Approaches to biosignature evaluation emphasising minimising false positives have been discussed previously in the Astrobiology literature. These mostly focus on searching for `anomalies'  \citep{cleland2019moving, cleland2019quest, kinney2022epistemology, smith2023life} that is, signals which have no abiotic explanation and virtually no chance of being false positives. However many of the anomalies proposed, such as sedimentary structures created by microbes or banded iron formations \citep{cleland2019quest} are not detectable on exoplanets by any planned or proposed observational techniques. For exoplanets we must consider biosignatures with a higher false positive probability. 

So, informed by the anomalies literature and the history and debate about false claims of alien life, we shift the focus from the usual object of Bayesian analysis, the probability of life given the evidence, to the probability of a false positive. This is also closer to the `usual' mode of science in focussing on falsification rather than verification \citep{popper2005logic}. However the intent of this paper is not to engage in the philosophy of science (for this see \citep{bayarri2004, kinney2022epistemology}). Our aim instead is to use the mathematical formalism of probability theory to discuss simple and general test cases for biosignatures. These `philosophical' considerations lead to a new kind of biosignature and highlight the strengths of different observational and modelling strategies. Actually carrying out these strategies will undoubtedly be a protracted and expensive endeavour and thus there is much value in taking a high level view about what different approaches can, even in principle, accomplish.

This paper is laid out as follows. In Section \ref{sec:framework} we introduce a simple probability theory formalism for biosignatures that highlights the trade-off between increasing the number of true positives (biosignature detected and life present) and reducing the number of false positives (biosignature detected and no life) arguing that the latter is much more important. We justify the neglect of false negatives (no biosignature detected and life present) in Section \ref{sec:negatives} and extend the formalism to multiple biosignatures in Section \ref{sec:multiple}. This allows us to formalise the previously proposed Confidence of Life Detection (CoLD) scale \citep{green2021call} and use it to discuss the detection of phosphine on Venus in Section \ref{sec:cold}. In Section \ref{sec:context} we discuss planetary context, for example, the distance from the host star, motivating the idea of a `peribiosignature', a biosignature that is stronger on the edges of a given parameter space. Gaia theory suggests one such example is habitability itself which we discuss in Section \ref{sec:time}, providing a way to mathematically express the Inhabitance Paradox/Gaian Bottleneck \citep{goldblatt2016inhabitance, chopra2016case}. In Section \ref{sec:obs} we compare the Bayesian and false positive approach and discuss observational strategy in light of some general arguments about the parameter dependence of the relevant probabilities. Finally, we summarise and discuss our findings in Section \ref{sec:discussion} and give some suggestions for future work.
 
\section{Probabilistic Framework}\label{sec:framework}
In this section, we use basic probability theory to formalise biosignature detection, before discussing false negatives and multiple biosignatures and finally how these considerations formalise previously proposed models for evidence assessment. 

\begin{figure}
    \centering
    \includegraphics[width=\linewidth]{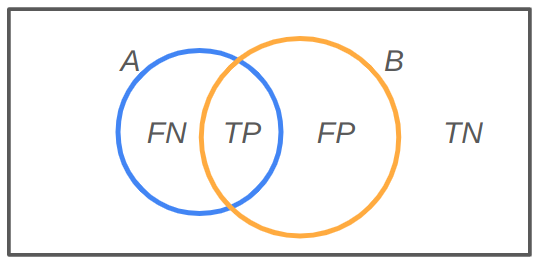}
    \caption{Events $A$ and $B$ are sets in the space of outcomes (the outer box). True/false positives/negatives are a partition of the sample space.}
    \label{fig:fp1}
\end{figure}

We start with a set $S$ of planets to survey. We will have a lot to say about planetary context \citep{catlingdavid2018exoplanet} in later sections, but to start assume that the planets are all similar enough to share a context e.g. Earth mass planets orbiting in the habitable zone of G-type stars. Let $A$ (aliens) be the set of planets, usually called an `event' in probability theory, where extraterrestrial life exists and $\bar{A}$ is the set where life does not exist. Trivially,
$$
|A| + |\bar{A}| = |S|,
$$
where $|A|$ is the size of the set $A$. We can divide through by $|S|$ to calculate the probabilities which must sum to unity,
$$
P(A) + P(\bar{A}) = 1,
$$
i.e. life exists or it doesn't.

Let $B$ be the event (i.e. set of planets) where a biosignature is observed and $\bar{B}$ the event where the biosignature is not observed, so
$$
P(B) + P(\bar{B}) = 1.
$$
The probability that life exists can be expressed in terms of $B$ as
$$
P(A) = P(A \cap B) + P(A \cap \bar{B}),
$$
i.e. life exists and is detected or life exists and is not detected. Similarly
$$
P(\bar{A}) = P(\bar{A} \cap B) + P(\bar{A} \cap \bar{B}).
$$
This shows there are four exclusive and exhaustive outcomes, illustrated in Figure \ref{fig:fp1}, namely,
\begin{enumerate}
    \item A True Positive,  $TP = P(A \cap B)$. Life exists and a biosignature is observed.
    \item A False Negative, $FN = P(A \cap \bar{B})$. Life exists and a biosignature is not observed.
    \item A False Positive, $FP = P(\bar{A} \cap B)$. Life does not exist and a biosignature is observed.
    \item A True Negative, $TN = P(\bar{A} \cap \bar{B})$. Life does not exist and a biosignature is not observed.
\end{enumerate}
Note, we are using the symbols $TP, FN$ etc. for probabilities, which will simplify the notation, but these symbols are sometimes used for counts, which can be obtained by multiplying the corresponding probability by $|S|$. 

We can link this to the Bayesian framework \citep[]{spiegel2012bayesian, catlingdavid2018exoplanet, walker2018exoplanet} by first writing down Bayes' theorem
$$
P(A|B) = \frac{P(A)P(B|A)}{P(B)},
$$
which relates the \emph{prior} probability of life $P(A)$ and the \emph{likelihood} of the biosignature $P(B|A)$ to the \emph{posterior} probability $P(A|B)$. This relates something we know or can estimate, $P(B|A)$, the probability of a biosignature given life, to the thing we want to know, $P(A|B)$, the probability of life, given the biosignature.

Using two elementary facts from probability theory, we can rewrite this in a couple of useful ways. First, the definition of conditional probability gives
$$
P(A \cap B) = P(A) P(B | A ).
$$
We also have
$$
P(B) = P(B \cap A) + P(B \cap \bar{A}),
$$
substituting these expressions into Bayes' theorem and using the definitions above gives
$$
P(A|B) = \frac{P(A)P(B|A)}{P(A)P(B|A) + P(\bar{A})P(B|\bar{A})} = \frac{TP}{TP + FP}.
$$
The first form is the one most commonly used to evaluate the posterior, the second form is often called the \emph{precision}, especially in the machine learning literature \citep{buckland1994relationship}. 

With the context fixed, we are only free to choose $B$ the biosignature. We can use this choice to optimise the precision in one of two ways: making $TP$ large or making $FP$ small. In Bayesian terms, by choosing $B$, we can control 
\begin{itemize}
\item $P(B|A)$ the true positive rate, also called \emph{sensitivity}.
\item $P(B|\bar{A})$ the false positive rate. The complement of this, $P(\bar{B}|\bar{A}) = 1 - P(B|\bar{A})$ is often called \emph{specificity}.
\end{itemize}
Avoiding false positives is an argument for using specific biosignatures rather than sensitive ones. 
\begin{itemize}
\item Consider a biosignature which is a radio signal encoding the first 100 binary digits of $\pi$. We do not know of any abiotic explanations for such signals, though they seem to be rare \citep{tarter2001search}. Thus an extraterrestrial radio signal would be a very specific biosignature (if we see it we are sure there is life) but not a sensitive one (we will miss lots of life which doesn't send such signals).
\item Let the biosignature, $B$, be atmospheric oxygen. It is plausible that complex life requires oxygen, but we also know of plausible abiotic mechanisms which can accumulate oxygen in the atmosphere \citep{krissansen2021oxygen}. Oxygen therefore seems to be a sensitive but not specific biosignature. 
\end{itemize}

\begin{figure}
    \centering
    \includegraphics[width=\linewidth]{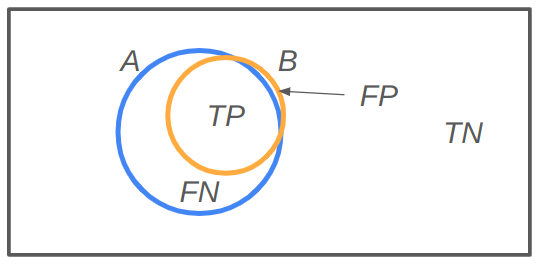}
    \caption{Events $A$ and $B$ where the false positive probability is very small.}
    \label{fig:fp1a}
\end{figure}
Minimising false positives is a probabilistic version of proof by contradiction, we accept an explanation $A$ if every alternative is significantly less likely than $A$. If the probability of a false positive $FP = P(B \cap \bar{A})$ is very small then
$$
P(B) = P(B \cap A) + P(B \cap \bar{A}) \simeq P(B \cap A).
$$ 
Figure \ref{fig:fp1a}, illustrates that if the false positive probability is small enough, $B$ is almost a subset of $A$ and seeing $B$ implies seeing $A$ (aliens). To state it in a more Bayesian way, 
$$
P(B) \simeq P(B \cap A) = P(B) P(A|B) \implies P(A|B) \simeq 1,
$$
the probability of life given the biosignature, $B$, is very high. This is in line with the prevailing `hypothetico-deductive' scientific method, which seeks to \emph{disprove} hypotheses i.e. show $P(B | \bar{A})$ is small, so the biosignature is inconsistent with the hypothesis of no life. On the other hand, if life implies a certain biosignature, or even guarantees it, it does not follow that the biosignature implies life. 

Clearly, we can have biosignatures which are too specific e.g. directed radio transmissions or particular complex molecules in the atmosphere. One way to have a low number of false positives is to have no positive detections! Thus, there is a trade-off between sensitivity and specificity, between biosignatures which can be detected versus potential false positives. We emphasise minimising false positives over maximising detections based on the following assumptions/observations:
\begin{enumerate}
\item The first goal is to discover at least one inhabited planet.
    \item There are an extremely large number of exoplanets \citep[e.g.][]{fressin2013,christiansen2022five}, so there are plenty of candidates to check for any biosignature. Although the sample of planets which are observable for a given biosignature may, currently, be very small (for example, searching for oxygen in Earth analogues), we are evaluating a framework agnostic to the precise biosignature and method of detection, so assume the sample size is large. 
    \item We have a fairly optimistic view on the prevalence of life in the universe, based mostly on how rapidly it appeared on Earth \citep{nisbet2001,kipping2025strong}.
    \item Previous and current debates over the interpretation of detected biosignatures, as briefly discussed in Section \ref{sec:introduction}, are usually disputed by proposing abiotic explanations for the observations. 
    \item There are relatively weak constraints on what alien life must do \citep{bartlett2020defining} and even defining what life is seems to be intractably difficult \citep{bich2018defining}. So there are few constraints on $P(B|A)$.
    \item On the other hand, while very difficult \citep{vickers2023confidence}, $P(B | \bar{A})$, the probability of a biosignature on a lifeless planet, can be approached via modelling and simulation based on universal physical principles.
    \item Even one confirmation of extraterrestrial life will be hugely informative for further searches.
    \item Given that the resources required to develop and maintain state--of--the--art observational facilities is extremely large, a small number of strong candidates might be preferable to a large number of weak ones.
\end{enumerate}
(i) Establishes our aim. We are not doing a survey of life in the galaxy, rather trying to find a single example of it not from Earth. (ii) and (iii) suggest we are not searching for a needle in a haystack. Thus even if we have quite specific biosignatures we can still find positive observations by checking many candidates. (iv-vi) are related to how confident we can be that what we do find is life. (vii) and (viii) recognise that while we have just one example of an inhabited planet it is impossible to know what is a particular characteristic of Earth and what is a feature of any inhabited planet. Confirming any life elsewhere immediately helps us to  narrow down generic features of inhabited planets, to direct further searches and clarify ambiguous or overlooked cases.

One interesting way to explore the trade-off between specificity and sensitivity is as a function of planetary context, $C$, like the type of host star, distance from the star, planet mass, etc. \citep{catlingdavid2018exoplanet}. Here it is important to remember that we care about the false positive \emph{probability} $FP(C) = P(B \cap \bar{A} | C)$ not just the false positive \emph{rate}. The confusion of the two ideas is called the `base rate fallacy' \citep{meehl1955antecedent}. Probability and rate are related via
$$
FP(C) = P(B \cap \bar{A} | C) = P(\bar{A} | C) P(B | \bar{A} , C).
$$
The number of false positives in a given context depends on both the false positive rate \emph{and} the prior, both of which vary with context, not necessarily in the same way. We will discuss this in Sections \ref{sec:context} and \ref{sec:obs} but for the remainder of this section we justify neglecting false negatives, discuss how to combine multiple biosignatures and finally show how to use this probability framework to formalise previously proposed criteria for life detection experiments.

\subsection{False Negatives}\label{sec:negatives}
Often considered in tandem with precision is \emph{recall}, defined as
$$
\frac{TP}{TP + FN},
$$
where, usually, increasing one decreases the other \citep{buckland1994relationship}. Just as precision is determined by the false positives, recall is controlled by the false negatives, cases where life exists but we fail to detect it. In this section we briefly discuss the different potential types of false negative in the context of planetary biosignatures.  Cases where the chosen biosignature is too specific, the case of "passenger life" where there is no significant impact on the wider planetary atmosphere, and detection of antibiosignatures which could falsely indicate the absence of life. 

\subsubsection{Life as we know it}
Firstly, we note that not every inhabited planet will produce a given biosignature. The Earth was certainly inhabited when there was little to no oxygen in the atmosphere, or vegetation, or night-time artificial light! If a biosignature is too specific, say a particular chemical known on Earth like chlorophyll, failure to detect it is only evidence that a particular kind of life is absent. Thus, strictly speaking, the event that we don't observe a particular biosignature, $\bar{B}$, is only evidence of the absence of life which produces $B$ \citep[and also assumes our observations have no measurement error, which we don't consider in this work for simplicity,][]{walker2018exoplanet}. For specific biosignatures e.g. oxygen, we should interpret evidence as for or against `oxygen producing life' rather than `life'. The desire to capture all kinds of life has motivated the development of agnostic or process-based biosignatures \citep{wong2024process} such as departures from chemical equilibrium in the planetary atmosphere \citep{hitchcock1967life, krissansen2018disequilibrium}, complexity of chemical networks \citep{sole2004large, wong2023toward}, complexity of reflected light signals \citep{bartlett2022assessing} or latent space representation of spectra \citep{cleaves2023robust}. By targeting life \emph{in general} these biosignatures aim to minimise false negatives.

\subsubsection{Passenger life}
More broadly, an inhabited planet may not produce any large-scale biosignature at all. Life as we observe it on Earth is a planetary phenomenon, deeply intertwined with the abiotic environment, Gaia \citep{lovelock1974atmospheric}. Life which is only a passenger on its host planet, i.e. which either does not interact significantly with the planetary environment, or does so on an extremely small or local scale, would be much more difficult, or even impossible to detect. \cite{kaltenegger2020high} suggests that Earth's atmosphere would have shown spectral features consistent with life since about 2 billion years ago, just over half of the time life has existed on Earth \citep{nisbet2001}. The search for alien life in our own solar system is the search for this kind of passenger life, and is extremely challenging even with (relatively) small distances and the possibility of landing on the surface to make in-situ observations \citep{styczinski2024chapter}. For exoplanets the situation is much worse. It is plausible that numerous exoplanets host some sort of passenger life, or even life which interacts strongly with its planet but produces only ambiguous biosignatures. 

\newcommand\BB{\reflectbox{B}}
\subsubsection{Antibiosignatures}

An antibiosignature is an indicator that a planet is not inhabited. These are more rarely discussed than biosignatures, with most attention given to CO \citep{kasting2014}. However, as with biosignatures which can be generated via abiotic pathways, CO can also accumulate through biotic processes \citep{schwieterman2019rethinking}. Other planetary observables, such as extremely high or low temperatures or an atmosphere in chemical equilibrium could also serve as antibiosignatures \citep{wogan2020chemical}. Let $\BB$ denote an antibiosignature which we would use to rule out a planet being inhabited. Another type of false negative is the observation of an antibiosignature on an inhabited planet. As before, we can express this in the language of probability 
$$
P(\BB \cap A) = P(A) P(\BB | A).
$$
Given an ambiguous candidate planet, identifying potential antibiosignatures could be important for refuting inhabitance. This is arguably the case for Venus. Despite the observation of the biosignature phosphine, the strong anti-biosignatures of an extreme surface temperature and harsh chemical environment still suggests an uninhabited planet, demanding further, in--situ observations \citep[e.g., The Venus Life Finder mission,][]{seager2021}.\\ 

The false negative rate for life detection experiments on exoplanets will likely be high. However, even one other confirmed example of an inhabited planet will vastly improve our understanding of life and give us better search criteria for finding these previously missed signals. We argue therefore that, given there are a lot of exoplanets and life is not vanishingly rare (we know it occurred at least once), the problem of false negatives on exoplanets is second-order, and we can tolerate a high number of false negatives until convincing evidence of any extraterrestrial life is found. If wide-ranging searches return nothing, this changes the assumptions and life (or the kind of life we can detect) may be  rare, in which case strategies to reduce false negatives will be necessary. 

Things are very different for life detection in the solar system, where observational prospects, however difficult, are much better. Again, even one example of life not on Earth would vastly improve our ability to find it elsewhere, so we should be extremely thorough in avoiding false negatives in the solar system.

\subsection{Multiple Biosignatures}\label{sec:multiple}
Thus far we have focused on the observation or potential identification of a single biosignature, however, our confidence that a planet is inhabited would be increased by observation of multiple biosignatures. Combining multiple biosignatures has often been discussed in the context of oxygen \citep{meadows2018exoplanet} and also recently in the context of phosphine detection on Venus \citep{cleland2022ammonia}.

\begin{figure}
    \centering
    \includegraphics[width=\linewidth]{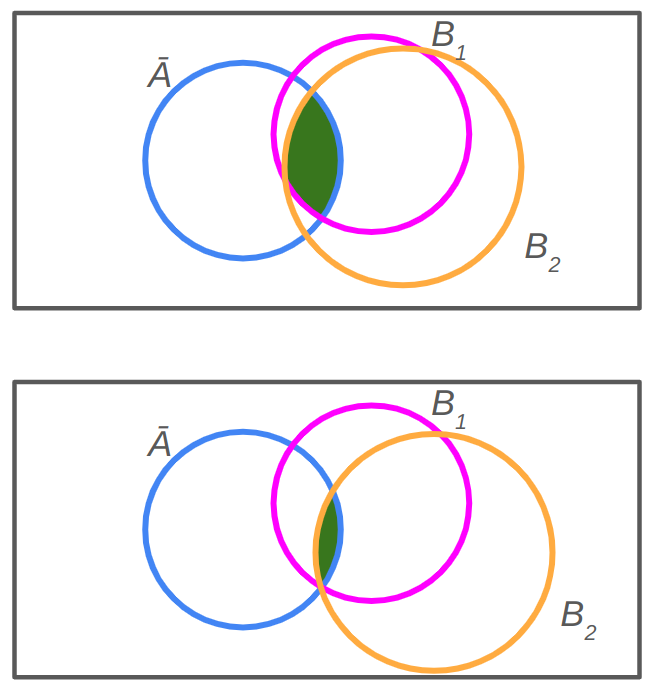}
    \caption{Top: Two biosignatures $B_1, B_2$ with large overlap. Detecting both reduces the probability of false positives (green shaded area) by a small amount. Bottom: Two biosignatures with less overlap. Detecting both reduces the probability of false positives (green shaded area) by a larger amount.}
    \label{fig:fp2}
\end{figure}
Consider two biosignatures $B_1$ and $B_2$ \footnote{These arguments can be extended to 3 or more in a straightforward but notationally more complex way}. The false positive rate is
$$
P(B_1 \cap B_2 | \bar{A}) = P(B_1 | \bar{A} ) P(B_2 | B_1, \bar{A} ),
$$
where $P(B_2 | B_1,  \bar{A} )$ is the probability of biosignature $B_2$ given no life and the observation of biosignature $B_1$.  As $P(B_2 | B_1, \bar{A} )$  is always less than one, more evidence can only decrease the false positive rate. On the other hand, requiring two different observations will also decrease the number of candidates and hence $TP$. As argued before, this trade off will generally be worth it, since we have many potential exoplanets to evaluate and the goal is to first achieve one positive confirmation.

The similarity of the biosignatures $B_1$ and $B_2$ determines how much adding the second observation reduces our uncertainty; the more similar they are the less the uncertainty is reduced, illustrated in Figure \ref{fig:fp2}. For example, if $B_1$ is oxygen and $B_2$ is ozone then $P(B_2 | B_1, \bar{A} )$ is relatively large - given oxygen, ozone is not surprising. On the other hand if $B_1$ is oxygen and $B_2$ is a large complex polymer, these are relatively independent $P(B_2 | B_1, \bar{A} ) \sim  P(B_2 |  \bar{A} )$ so the simultaneous observation of both is stronger evidence than either one independently. For more detailed and extensive analysis of multiple biosignatures see \citet{sandora2020biosignature, fields2023information}. In this work, for simplicity,  we mostly focus on a single biosignature. 

\subsection{The CoLD Scale}\label{sec:cold}

\cite{green2021call} propose the Confidence of Life Detection (CoLD) scale for evaluating and communicating claims of extraterrestrial life detection. This is a progressive series of levels from 1 (lowest level of evidence) to 7 (confirmation). It involves two separate steps of false positive assessment, which involves ruling out all \emph{prior} abiotic explanations for a biosignature and then ruling out abiotic explanations proposed \emph{after} the initial discovery. Although there are alternatives \citep{ neveu2018ladder, meadows2022community, vickers2023confidence} the CoLD system is particularly apt for formalising in a probability framework and highlights the importance of false positive reduction.

One should keep in mind that this framework is designed for evaluating claims about life on a single planet, whereas up to now we have been discussing large collections of planets. This poses some challenges for interpreting the meaning of probabilities e.g. as expressing subjective belief or averages over alternative histories of the same planet, see \citep{bayarri2004, kinney2022epistemology} for more on this.

\begin{table*}
    \centering
    \resizebox{\textwidth}{!}{\begin{tabular}{|c|c|c|}
    \hline 
    Level     &  Description & Probability Framework\\ \hline \hline 
    1   & Detection of a signal known to result from biological activity & The event $B \cap \epsilon$, where $\epsilon$ is measurement error\\ \hline 
    2   & Contamination ruled out & $B \cap \epsilon \simeq B$\\ \hline 
    3   & Demonstration or prediction of biological production of signal in the environment of detection & $P(B|A,C) \neq 0$ \\ \hline 
    4   & All known non-biological sources of signal shown to be implausible in that environment & $P(B \cap \bar{A}|C) < \alpha_1$\\ \hline 
    5   & Additional independent signal from biology detected & The event $B_2$ where $P(B_2|B) \simeq P(B_2)$, giving $P(B \cap B_2 \cap \bar{A | C}) < P(B \cap \bar{A}|C)$\\ \hline 
    6   & Future observations that rule out alternative hypotheses proposed after the original announcement & $P(B \cap B_2 \cap \ldots \cap B_n \cap \bar{A}|C) < \alpha_2 \ll \alpha_1$\\ \hline 
    7   & Independent follow-up observations of predicted biological behaviour in the environment & The event $B_{pred}$ where $P(B_{pred} | A, C) \simeq 1$\\ \hline 
    \end{tabular}}
    \caption{The Confidence of Life Detection (CoLD) scale from \citep{green2021call}. We have added the third column showing how to express the various levels in the probabilistic language used in the text. }
    \label{tab:cold}
\end{table*}

Table \ref{tab:cold} shows the CoLD scale, where the description given in \cite{green2021call} is in the middle column and our mathematical formulation in the right column. \cite{green2021call} use the example of the Martian meteorite ALH84001 to demonstrate the scale, we can do a similar analysis for Venus and phosphine. $B$ is the observation of phosphine and $A$ is (phosphine producing) life. The original detection by \cite{greaves2021phosphine} would reach Level 1. While there is some dispute about the legitimacy of the detection \citep{villanueva2021no, greaves2021reply}, it seems that follow up observations do support the presence of phosphine \citep{clements2024venus}, thus we can cautiously pass Level 2. We note in passing that, since the CoLD scale was also designed for communicating about life detection experiments, this level is a potential source of confusion for lay audiences, where the confidence that $B$ is observed i.e. $\epsilon$ is small, is confused with the probability of life detection $P(A|B)$. If the probability of experimental error is quoted e.g. \citep{madhu2025} great care should be taken to clarify that this is not a statement about the confidence that life has been detected.

Returning to the scale, phosphine is known to be produced by life on Earth, however the context of Venus is quite different. \cite{bains2024venus} discuss some plausible mechanisms for life to survive and produce phosphine in the Venusian atmosphere. To pass Level 3 we only need to show that life could exist, not how likely it is, so this is achieved. Level 4 is the first stage where we must rule out false positives. \cite{bains2024source} discuss ruling out a large number of abiotic sources of phosphine (geothermal processes, atmospheric chemistry etc.). In our language this would be putting bounds on $P(B|\bar{A})$ and showing it is small. Based on known properties of Venus (high temperature, lack of water) our prior expectation would be that life on Venus is quite unlikely and Venus is outside most common definitions of the solar Habitable Zone \citep{ramirez2018more}. Hence we can assume, for Venus, $P(\bar{A}) \simeq 1$ and thus $P(B \cap \bar{A}) \simeq P(B|\bar{A})$. While we cannot quantify it numerically, showing that there are few if any plausible abiotic mechanisms to generate phosphine, as done by \cite{bains2024source}, arguably passes Level 4.

Level 5 requires an independent follow up signal e.g. some other chemical by-product of the putative Venusian, phosphine-generating life. This level has not been passed, to do so in--situ samplings from spacecraft have been proposed \citep{bains2024venus}, which would also serve as a way to fully exclude the possibility of measurement error. Apart from spacecraft missions, this kind of analysis suggests at this stage turning the focus from phosphine towards some other signals which could be produced by Venusian life. Progress might be made with modelling and theoretical efforts targeted towards  proposing new candidate biosignatures, $B_2$, ideally also detectable in spectra and, following Section \ref{sec:multiple}, independent of phosphine. Some work has been done here looking at ammonia as a biosignature \citep{cleland2022ammonia}.

Level 6 is a more stringent test for False Positives which would require ruling out, with high confidence, any abiotic explanations for the biosignatures detected at the previous level. Since we are evaluating a single planet, statistical quantification seems unlikely or at least would be somewhat contrived. Even if possible, different fields have different standards of evidence \citep{benjamin2018redefine} with e.g. a one in a million chance of error required for discovering a new elementary particle \citep{higgs}.  It is more likely that some kind of community consensus will be reached after accumulation of sufficient evidence and failure of abiotic models to account for it \citep{bains2021astrobiologists}. 

Level 7 would require a model of how Venusian life produces the observed biosignatures. Given the ingeniousness of modellers and the flexibility of life, this seems likely to always be achieved. After that we could confidently say life on Venus had been detected and Level 7 represents the first steps towards studying it. Note the role played by False Positive detection (4 and 6) versus confirmatory prediction (3 and 7) in this framework. Step 3 is not strongly constrained, all kinds of exotic biology e.g. non-carbon based life, could be proposed to give a non-zero probability for some form of life to generate a signal. Levels 4, 5 and 6 are the key steps in building confidence that the observed data does not have some alternative explanation. Level 7, showing how life can generate the observed data, is the cherry on the cake rather than the foundation of our belief. 

\section{Peribiosignatures}\label{sec:context}

As mentioned, biosignatures depend on the \emph{planetary context} \citep{chan2019deciphering, catlingdavid2018exoplanet}. Take for example $B$, liquid water on the surface of a planet and $r$ the distance of this planet from its host star. We use this simple case for building intuition, but the same reasoning will hold for other biosignatures (departures from atmospheric chemical equilibrium, presence of various gases etc.) considered relative to other planetary contexts (age, size of planet, type of host star etc.). The false positive probability depends on the context parameter $r$ and can be expressed as
$$
  FP(r) = P(B \cap \bar{A} | r) = P(\bar{A} | r) P(B | \bar{A}, r),
$$
where $P(\bar{A} | r)$ is the background rate e.g. the probability to observe (no) life on a planet at parameter value $r$.  $P(B | \bar{A}, r)$ is the probability of biosignature $B$, given no life, at parameter value $r$. 

For the case where $r$ is the distance of the planet from the host star $P(\bar{A} | r)$ fairly directly represents the Habitable Zone (HZ), or more precisely, the abiotic HZ where planetary properties are computed in the absence of feedbacks and interactions between life and its host planet. $P(\bar{A} | r)$ will depend on factors such as the type of host star, and planetary atmospheric composition, with extensive work in the literature focused on exploring this concept through abiotic climate simulations \citep[for example,][]{turbet2019,turbet2023}. Fixing all other variables, we would expect $P(\bar{A} | r)$ to be high very close to or far from the star, and to be smaller in a region where temperatures allow liquid water to exist \citep{ramirez2018more}.

Allowing that biosignatures depend on context allows us to widen the scope of what can be considered a biosignature e.g. for liquid water, within the abiotic HZ $P(B | \bar{A}, r)$ is, by definition, large, so water is a weak biosignature in the HZ. Outside and on the edge of the abiotic HZ, $P(B | \bar{A}, r)$ is much smaller and the observation of liquid water would imply something is keeping it liquid. One such mechanism is life regulating the global temperature. 

We define a \textbf{peribiosignature} (`peri-' meaning edge) as any substance, group of substances, or phenomenon that provides evidence of life \emph{where life would a priori be unlikely}. We have in mind examples like liquid water outside or on the edge of the abiotic HZ. In Earth history, when the HZ was closer to the star, this `peribiosignature' is referred to as the faint young sun paradox \citep{sagan1972earth}, which life is argued to play a large role in resolving \citep{charnay2020faint}. Similarly, temperatures which were hotter or colder than expected, measured e.g. via thermal emission \cite{greene2023thermal}, would be evidence that something, like life, is regulating the temperature, with numerous models \citep{watson1983biological, arthur2023gaian} suggesting that life tends to regulate temperature in its favour.

The broader requirement, in terms of exoplanet biosignatures, for life to impact its planetary environment, i.e. not "passenger life", is closely related to Gaia theory. Although the concept of Gaia has been perpetually controversial, the basic fact that life on Earth couples with the abiotic environment in a deep and generally beneficial way is not really disputed \citep{lenton2018selection, arthur2022selection, wong2024process}. Our own modelling studies \citep{arthur2023gaian, nicholson2023biotic} and older ones \citep{schwartzman1991geophysiology} suggest Gaia, the maintenance of beneficial conditions by and for life, implies that the Habitable Zone could be `wider' than physical and chemical considerations alone would imply. If life is able to regulate the environment it could keep conditions habitable despite external forcing, an observation with a long history in Gaia theory \citep{watson1983biological}.

The concepts of inhabitance and habitability have been noted before as being inextricably linked, a concept dubbed the `Inhabitance Paradox' \citep{goldblatt2016inhabitance}. This idea is that different planetary climate states, from global glaciation to a runaway greenhouse can be stable in the same planetary context. One of the primary ways Earth avoids runaway scenarios is regulation of the atmosphere by life. Thus for a planet to remain habitable for long time-spans it must be inhabited. Related is the Gaian Bottleneck \citep{chopra2016case}, the idea that when life emerges it must quickly evolve to regulate the planetary atmosphere, to, among other things retain water on the surface \citep{harding2010water}, something which seems not to have occurred for the apparently lifeless and notably less wet planets Venus and Mars. These considerations suggest habitability itself is a peribiosignature.

\begin{figure}
    \centering
    \includegraphics[width=\linewidth]{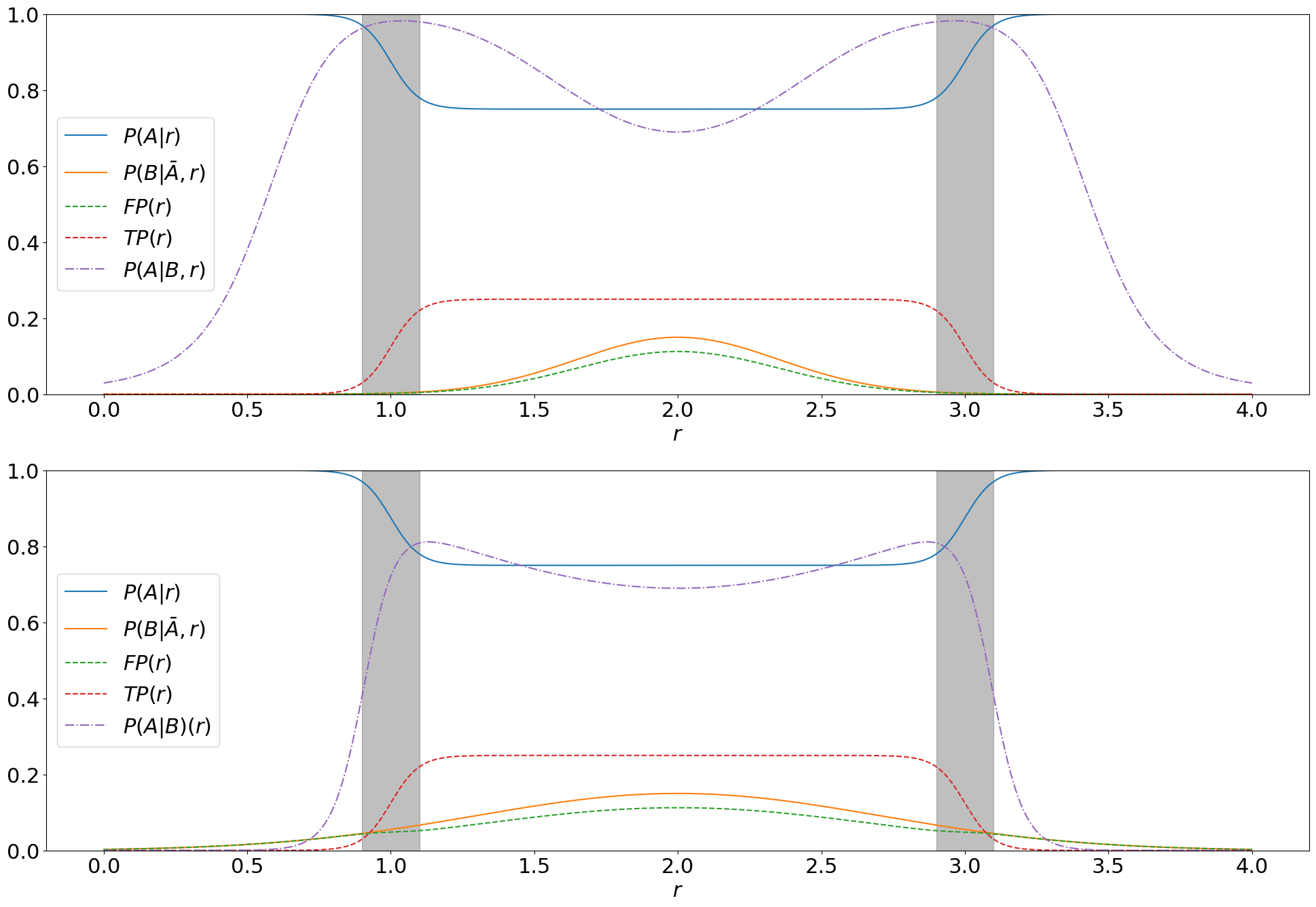}
    \caption{Illustrative probability functions used to calculate the posterior/precision. The grey bands give a region where $B$ is a good (peri)biosignature. Top shows a narrow profile for false positive rate, bottom shows a broader profile.}
    \label{fig:pb1}
\end{figure}
 Figure \ref{fig:pb1} shows a simple illustration of a peribiosignature. The abiotic habitable zone is defined by  $P(\bar{A} | r)$, drawn as a region, $1 \lessapprox r \lessapprox 3$, where the probability for life is small but non-zero, $0.25$, and is almost 0 elsewhere. To model a peribiosignature, like habitability, we use a Gaussian centred in the middle of the HZ for $P(B|\bar{A}, r)$. By inhabitance/bottleneck type arguments we set $P(B|A) = 1$. By definition, habitability is common on abiotic worlds in the HZ, so here it is a bad biosignature with a high number of False Positives. However, towards the edges (grey bands in Figure \ref{fig:pb1}) there are regions where the False Positive rate is low, and the probability of life is non-zero, making observations of a habitable conditions in this zone themselves a strong indication of life. 
 
 Note, in the bottom panel of Figure \ref{fig:pb1}, we have $P(A|B, r=2) \simeq P(A|B, r=1)$, that is, the posterior/precision is equal in the centre of the HZ and the edge. However the false positive rate and probability are quite different, by a factor of about $2$. In the centre of the HZ we have more detections, at the edge fewer false positives.  Precision only depends on $\frac{FP}{TP}$, so an increase in one can compensate for an increase in the other or \emph{vice versa}. 

\subsection{The Inhabitance Paradox and the Hazard Function}\label{sec:time}
It is more complex to incorporate time as planetary context but this will be crucial for understanding complex biosignatures. For an inhabited planet let $T$ be the time until life goes extinct. The cumulative distribution function
$$
F(t) = P(T \leq t),
$$
gives the probability of extinction before time $t$ and the \emph{survivor function} 
$$
S(t) = 1 - F(t) = P(T > t),
$$
then gives the probability of life surviving past time $t$. The probability density, 
$$
f(t) = \frac{dF(t)}{dt} = -\frac{dS(t)}{dt} = \lim_{dt \rightarrow 0} \frac{P(t \leq T \leq t+dt)}{dt}
$$
is the instantaneous probability of extinction i.e. the probability of surviving exactly up to time $t$ and then going extinct in the infinitesimal interval $(t,t+dt)$. The conditional probability\footnote{Using $P(A \cap B) = P(A) P(B|A)$ again}, called the \emph{hazard function}, is given by
$$
h(t) = \frac{f(t)}{S(t)} \simeq \lim_{dt \rightarrow 0} P( \text{go extinct in } t+dt | \text {survived to } t).
$$

\begin{figure}
    \centering
    \includegraphics[width=\linewidth]{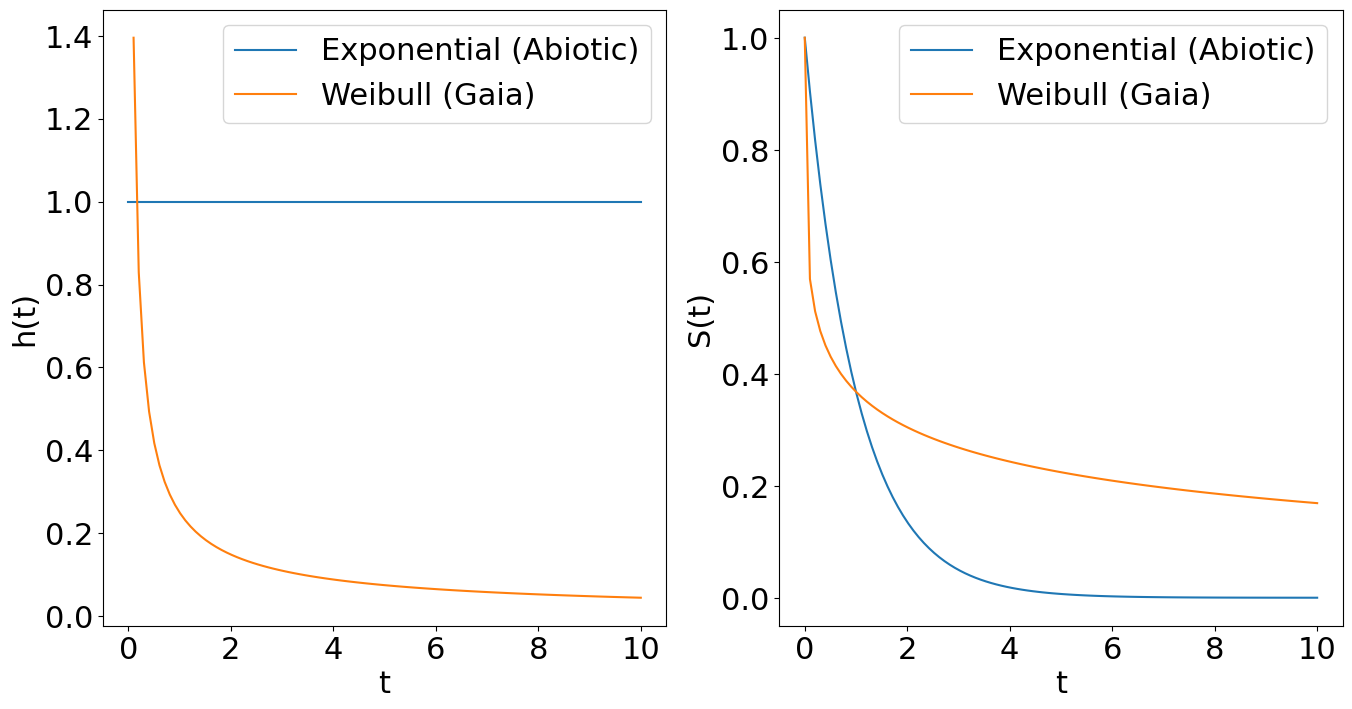}
    \caption{Left: illustration of a hazard function for memoryless (abiotic) and memoryful (Gaian) assumptions. Right: The corresponding survivor functions. The rate parameter for both is $\lambda=1$ and the shape parameter $p = 0.25$.}
    \label{fig:haz1}
\end{figure}
Assuming a static or very slowly changing stellar environment (i.e. not too early or late in the host star's life cycle), if inhabitance plays no role in habitability the hazard function is approximately independent of $t$, $h(t) = \lambda$. This is a memoryless process and results in an exponentially decreasing probability of survial, $S(t) = e^{-\lambda t}$. Gaia theory posits that life has a positive effect on habitability and therefore that the hazard function is \emph{decreasing} i.e. unlike machines or people, the longer Gaia persists the longer she is likely to persist. The Gaian bottleneck proposes that early on, $t\simeq 0$, the hazard function is large. This behaviour is often modelled through a Weibull distribution \citep{heckert2002handbook}. We can simply think of this as a convenient distribution whose hazard function, $h(t) = \lambda p (\lambda t)^{p-1}$, has the desired shape for $p<1$. Figure \ref{fig:haz1} shows the survivor functions for Gaian and non-Gaian assumptions showing that Gaia predicts a much larger probability of survival for inhabited planets than the abiotic model. Combined with the inhabitance paradox discussed above, a peribiosignature using time as a parameter would be a habitable older planet.

The properties of the host star of course also change over time, resulting in the need to examine a combined parameter space $FP(r,t)$. This paper only aims to provide a conceptual framework so we leave this more complex modelling, incorporating the real physics involved in HZ calculations, for the future work.

\section{Observing Biosignatures}\label{sec:obs}
As discussed in Section \ref{sec:framework}, \cite{catlingdavid2018exoplanet} and \cite{walker2018exoplanet} suggest using Bayes' theorem to analyse biosignatures. Factoring $TP$ from the \emph{precision} yields,
\begin{align*}
 P(A|B) = \frac{1}{1+\frac{FP}{TP}}.
\end{align*}
As opposed to CoLD, the posterior/precision balances avoiding false detections with increasing true ones.

Generally in Bayesian analysis, when the evidence is strong, the effect of the prior on the posterior probability, $P(A|B)$, is small. For something like a technosignature or a herd of elephants on Mars, $P(B|\bar{A})$ is practically 0 and hence $P(A|B) = 1$ regardless of $P(A)$. For biosignatures with abiotic pathways available to produce them this is not the case. To avoid the prior \cite{walker2018exoplanet} suggest using `detectability' instead of the posterior, defined as
$$
D = \frac{P(B|A)}{P(B|\bar{A})}.
$$
This is also known as the Bayes Factor and compares the likelihood of the evidence $B$ under two competing hypotheses: life or no life, with large values favouring life. 
 
Both detectability and the posterior depend on $P(B|A)$, the probability of a biosignature given life. By choosing biosignatures, $B$, and contexts, $r$ we can try to maximise the posterior and detectability. This is achieved with a large prior $P(A|r)$ and/or a low false positive rate, $P(B | \bar{A}, r)$. We can also try to increase $P(A|B)$ that is, look for data which is likely to be produced by life. For  biosignatures which are chosen \emph{because} they are produced by life $P(A|B) \simeq 1$ by design. `Process based' biosignatures like chemical disequilibrium are potentially of this type. In this case we again focus on minimising $FP$ e.g. by varying the context. In other cases, as discussed, the constraints on $P(B|A)$ are quite weak. Thus, while the Bayesian analysis apparently balances $FP$ and $TP$, as long as we avoid trivial cases (far outside the HZ, extremely specific biosignatures) where $TP = 0$, the most important factor is still $FP$, the probability of a false positive.

Obviously the exact dependence of the probabilities on the context is going to greatly affect the details but we can make some reasonable assumptions when for an example where $r$ is the distance from the host star:
\begin{enumerate}
    \item Abiotic models of the HZ \citep[e.g.][]{ramirez2018more} are a good starting point for $P(A | r)$. We and others have argued that life can widen the HZ, which means we might expect more life at the edges of the HZ than the abiotic model predicts. Conversely, since life can also cause deleterious effects (snowball earth, runaway greenhouse etc.) we might expect less life, or more precisely, fewer cases of habitable conditions, inside the HZ than expected from abiotic predictions.
    \item The true positive rate $P(B|A,r)$ should generally be much higher than $P(B|\bar{A},r)$ and probably fairly flat as a function of $r$. There seems to be little point in trying to search for a biosignature unlikely to be generated by life, that is very ambiguous or very sensitive to the planetary context.
    \item The false positive rate $P(B| \bar{A},r)$ should be reasonably small. 
    \item Because $P(B|A,r)$ always appears in combination with $P(A|r)$ its dependence on $r$ far outside the HZ, where $P(A|r) \simeq 0$, does not need to be estimated. 
\end{enumerate}

\begin{figure*}
    \centering
    \includegraphics[width=1\textwidth]{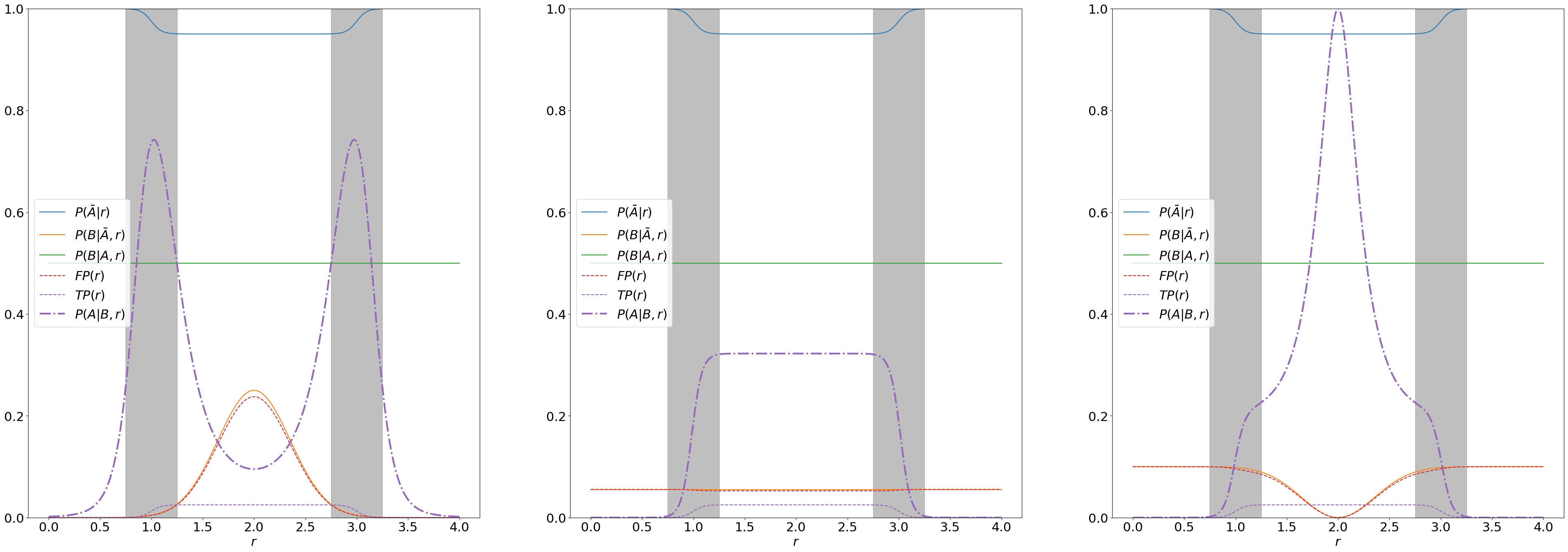}
    \caption{Left: Concave false positive rate. Centre: Flat false positive rate. Right: convex false positive rate.}
    \label{fig:fpexamples}
\end{figure*}
Figure \ref{fig:fpexamples} uses these assumptions and shows three idealised possibilities for false positive rates: concave, flat and convex dependence on $r$. In the latter case it is better to observe the centre of the HZ, in the first we would be better looking at the edges. If $P(A|r)$ is taken as the abiotic HZ we should also expect a boost in positive detections at the edge and a dip in the middle \citep{arthur2023gaian}, also suggesting that the edges are a better place to look in the flat case, i.e. if false positives are equally likely throughout the HZ for the specific biosignature.

Another, reasonable, assumption is that our ability to detect a biosignature is independent of whether it is produced by life or not. How well we can detect a particular biosignature, using particular instruments \citep{fujii2018exoplanet}, changes the number of potential candidates and the experimental error, but does not change the posterior probability. In the case of a biosignature with false positive dependence as in the left panel of Figure \ref{fig:fpexamples}, an observing strategy and instrumentation that yields better performance at the edges of the HZ is actually better for the detection of this type of biosignature. Currently envisaged techniques to detect biosignatures, such as transit spectroscopy \citep[e.g.][]{cadieux2024} and direct imaging \citep[e.g.][]{bowens2021} do, in a qualitative sense, show such a dependence on orbital distance. For transmission spectroscopy, planets orbiting closer to smaller/cooler host stars, compared to the Earth, are favoured given it is a measurement of the relative radii of the host star and planet that can be repeated once per orbital period, or year. Whereas, for direct imaging techniques, greater orbital separations are favoured, allowing easier separation of the planetary, and host star, flux. 

We concede that the specific performance of a given facility or instrument is a complex factor of many parameters. However, it is clear that modelling the dependence of false positive rate and prior on orbital distance and other parameters \citep[such as planetary mass, see, ][]{kopparapu2014habitable} will be highly informative. Our aim here is to illustrate a way of prioritising, or structuring our search for a detection of life, and demonstrate that the `centre' of the  HZ, or some narrow (in terms of orbital radii) definition of it, is not \emph{ipso facto} the most efficient strategy. Researchers can very usefully contribute to the planning of future observational campaigns by attempting to compute how $P(A|r)$ the prior for life and $P(B|\bar{A},r)$ the false positive rate depend on various parameters of the planetary system, such as the distance from the host star as in the example we have illustrated here.

\section{Discussion}\label{sec:discussion}

We know of just one example of an inhabited planet and have only a few hints to guide our search for alien life. One strategy is to focus on  biosignatures with no good abiotic explanation, rather than the physical and chemical signals which have a much higher potential for ambiguity. The anomalies of \cite{cleland2019quest} or even technosignatures indicative of intelligent life \citep{wright2022case, lingam2023technosignatures} would on their own provide the extraordinary evidence needed to making a convincing case for life. Absent these, more holistic measures like chemical disequilibrium or `process based biosignatures' \citep[\emph{Gaiasignatures},][]{wong2024process} may have less ambiguity and less specificity for life as we know it on Earth.

In this work we introduce the idea of a peribiosignature - a biosignature which is stronger on the edge of parameter space, for example on the edge of the abiotic habitable zone or at late times in stellar evolution. Gaia theory suggests that inhabitance and habitability are closely interlinked. Thus we propose habitability itself, measured by e.g. planetary temperature \citep{greene2023thermal} or the presence of liquid water as a peribiosignature in the parameter spaces of stellar distance and stellar evolution. Chemical disequilibrium in the atmosphere is another peribiosignature, abiotically likely early in a planet's history and not likely at late times \citep{krissansen2018disequilibrium}. Using atmospheric composition or radius as the parameter, highly extended, hydrogen dominated atmospheres on smaller planets in orbital configurations where hydrogen would be expected to have escaped is another potential peribiosignature (in this case of methanogenic life). Life on Earth today plays a large role in preventing hydrogen escape \citep{harding2010water, matassa2015resource}, though this was not always the case, with \citep{catling2001biogenic} suggesting that late times are also necessary to make this an effective peribiosignature.

We have contrasted the Bayesian posterior  \citep{catlingdavid2018exoplanet, walker2018exoplanet} with an approach that focusses mainly on false positives, in particular the CoLD framework \citep{green2021call}. Both approaches require modelling $P(B|\bar{A})$, the probability to abiotically generate a biosignature and $P(A)$ the prior probability of life. While extremely difficult \citep{vickers2023confidence} estimating $P(B|\bar{A})$ is at least constrained by known and relatively understood laws of physics, chemistry and geology. Despite the difficulty, there is really no other option, in any framework, than ruling out alternatives. Indeed, one motivation for picking certain biosignatures over others could be that $P(B|\bar{A})$ is small and computable. As for the prior, we highlighted in Section \ref{sec:context} that understanding the dependence of $P(A)$ on planetary parameters can be quite helpful, however more work is needed on life's effect on extending the HZ or modifying the HZ \citep{nicholson2023biotic, arthur2023gaian, arthur2024does}.

The Bayesian approach also requires estimating $P(B|A)$, the probability of observing a biosignature, given life. With the lack of constraints on alien biology and what it can feasibly do, this is an extremely difficult task. Focusing on general principles like punctuated equilibrium \citep{arthur2022selection} or nutrient limited growth \citep{nicholson2022predicting}, seems to be a productive way forward here.

 Different situations and biosignatures might favour different approaches. For example, we should cast a wide net for technosignatures since they appear to be rare but they are such strong evidence the false positive rate (absent experimental error) will be virtually zero \citep{tarter2001search}. Efforts at life detection in the solar system have only a very limited number of places to look, so need to thoroughly examine each one and false negatives are a more important concern here. However the situation with exoplanets is different. With 5000+ exoplanets already found and many more expected \citep{christiansen2022five}, and expensive missions concepts under consideration, we can perhaps afford to ignore some candidate planets and biosignatures to focus searches where ambiguity is lower. 

We also suggest that while there remains no confirmation of any extra-terrestrial life, collecting lots of plausible but ambiguous candidates is a recipe for generating false claims and heated arguments and, perhaps, wasted effort. While perhaps unintuitive, we believe the analysis offered here, which suggests that searching for signs of life in unlikely places, or looking for rarer but less ambiguous signals is more likely to lead to a consensus confirmation of extraterrestrial life than searching for, say oxygen in the middle of the habitable zone. When it comes to life $N=2 \gg N=1$ and after any detection is confirmed we can return to plausible but ambiguous candidates vastly more informed.

This paper presents a theoretical framework for thinking about biosignatures, and in particular false positives, allowing us to formalise scales like CoLD and provide tools for thinking about observational strategy. The hard work of detailed, physical, chemical and biological modelling to estimate these probabilities and rule in or out various signals of course remains. This work aims to productively guide these efforts and we offer the following suggestions
\begin{enumerate}
\item Search for unambiguous evidence, like technosignatures.
\item Accept a high false negative rate until the first detection of extraterrestrial life is made.
\item For biosignatures with abiotic explanations, focus searches on regions of parameter space (star type, distance from star, planetary size, age...) where abiotic mechanisms to generate the biosignature are less likely (peribiosignatures).
\item Habitability itself is a biosignature.
\item Where possible, search for multiple, independent biosignatures.
\item Modelling should focus on $P(B|\bar{A})$, by studying abiotic mechanisms that can generate biosignatures, and $P(A)$ by studying the probability for life to emerge \citep{foote2023false} and persist \citep{arthur2024does} as a function of planetary context. 
\item The design of observational instruments, facilities and strategies needs to combine experts from theory, observation and engineering to identify regions of parameter space with high priors \textbf{and} low false positive rates to focus resources there.
\end{enumerate}

At this moment in time we know that life is present on Earth (and has been for almost all of its `lifetime'), but do not know whether it exists elsewhere. Neither the hypothesis that life does exist elsewhere, nor the hypothesis that it is unique to Earth, can currently be rejected. We propose this new conceptual framework without making a judgment as to whether either hypothesis is more likely, and in recognition that both possibilities are immensely intriguing!

\section*{Acknowledgements}

This work was supported by a UKRI Future Leaders Fellowship [grant number MR/T040866/1], a Science and Technology Facilities Council Consolidated Grant [ST/R000395/1], a Science and Technology Facilities Council astronomy observation and theory small award [ST/Y00261X/1], and the Leverhulme Trust through a research project grant [RPG-2020-82].




\bibliographystyle{mnras}
\bibliography{edge} 

\begin{thebibliography}{}
\makeatletter
\relax
\def\mn@urlcharsother{\let\do\@makeother \do\$\do\&\do\#\do\^\do\_\do\%\do\~}
\def\mn@doi{\begingroup\mn@urlcharsother \@ifnextchar [ {\mn@doi@} {\mn@doi@[]}}
\def\mn@doi@[#1]#2{\def\@tempa{#1}\ifx\@tempa\@empty \href {http://dx.doi.org/#2} {doi:#2}\else \href {http://dx.doi.org/#2} {#1}\fi \endgroup}
\def\mn@eprint#1#2{\mn@eprint@#1:#2::\@nil}
\def\mn@eprint@arXiv#1{\href {http://arxiv.org/abs/#1} {{\tt arXiv:#1}}}
\def\mn@eprint@dblp#1{\href {http://dblp.uni-trier.de/rec/bibtex/#1.xml} {dblp:#1}}
\def\mn@eprint@#1:#2:#3:#4\@nil{\def\@tempa {#1}\def\@tempb {#2}\def\@tempc {#3}\ifx \@tempc \@empty \let \@tempc \@tempb \let \@tempb \@tempa \fi \ifx \@tempb \@empty \def\@tempb {arXiv}\fi \@ifundefined {mn@eprint@\@tempb}{\@tempb:\@tempc}{\expandafter \expandafter \csname mn@eprint@\@tempb\endcsname \expandafter{\@tempc}}}

\bibitem[\protect\citeauthoryear{Arthur \& Nicholson}{Arthur \& Nicholson}{2022}]{arthur2022selection}
Arthur R.,  Nicholson A.,  2022, Journal of Theoretical Biology, 533, 110940

\bibitem[\protect\citeauthoryear{Arthur \& Nicholson}{Arthur \& Nicholson}{2023}]{arthur2023gaian}
Arthur R.,  Nicholson A.,  2023, Monthly Notices of the Royal Astronomical Society, 521, 690

\bibitem[\protect\citeauthoryear{Arthur, Nicholson  \& Mayne}{Arthur et~al.}{2024}]{arthur2024does}
Arthur R.,  Nicholson A.,   Mayne N.,  2024, Monthly Notices of the Royal Astronomical Society, 533, 2379

\bibitem[\protect\citeauthoryear{Bains \& Petkowski}{Bains \& Petkowski}{2021}]{bains2021astrobiologists}
Bains W.,  Petkowski J.~J.,  2021, International Journal of Astrobiology, 20, 312

\bibitem[\protect\citeauthoryear{Bains et~al.,}{Bains et~al.}{2021}]{bains2021phosphine}
Bains W.,  et~al., 2021, Astrobiology, 21, 1277

\bibitem[\protect\citeauthoryear{Bains, Seager, Clements, Greaves, Rimmer  \& Petkowski}{Bains et~al.}{2024a}]{bains2024source}
Bains W.,  Seager S.,  Clements D.~L.,  Greaves J.~S.,  Rimmer P.~B.,   Petkowski J.~J.,  2024a, Frontiers in Astronomy and Space Sciences, 11, 1372057

\bibitem[\protect\citeauthoryear{Bains, Petkowski  \& Seager}{Bains et~al.}{2024b}]{bains2024venus}
Bains W.,  Petkowski J.~J.,   Seager S.,  2024b, Astrobiology, 24, 371

\bibitem[\protect\citeauthoryear{Bartlett \& Wong}{Bartlett \& Wong}{2020}]{bartlett2020defining}
Bartlett S.,  Wong M.~L.,  2020, Life, 10, 42

\bibitem[\protect\citeauthoryear{Bartlett et~al.,}{Bartlett et~al.}{2022}]{bartlett2022assessing}
Bartlett S.,  et~al., 2022, Nature Astronomy, 6, 387

\bibitem[\protect\citeauthoryear{Bayarri \& Berger}{Bayarri \& Berger}{2004}]{bayarri2004}
Bayarri M.~J.,  Berger J.~O.,  2004, Statistical Science, 19, 58

\bibitem[\protect\citeauthoryear{Benjamin et~al.,}{Benjamin et~al.}{2018}]{benjamin2018redefine}
Benjamin D.~J.,  et~al., 2018, Nature human behaviour, 2, 6

\bibitem[\protect\citeauthoryear{Bich \& Green}{Bich \& Green}{2018}]{bich2018defining}
Bich L.,  Green S.,  2018, Synthese, 195, 3919

\bibitem[\protect\citeauthoryear{{Bowens} et~al.,}{{Bowens} et~al.}{2021}]{bowens2021}
{Bowens} R.,  et~al., 2021, in European Planetary Science Congress. pp EPSC2021--79, \mn@doi{10.5194/epsc2021-79}

\bibitem[\protect\citeauthoryear{Buckland \& Gey}{Buckland \& Gey}{1994}]{buckland1994relationship}
Buckland M.,  Gey F.,  1994, Journal of the American society for information science, 45, 12

\bibitem[\protect\citeauthoryear{CERN}{CERN}{2025}]{higgs}
CERN 2025, Why do physicists mention “five sigma” in their results?, \url {https://home.cern/resources/faqs/five-sigma}

\bibitem[\protect\citeauthoryear{{Cadieux} et~al.,}{{Cadieux} et~al.}{2024}]{cadieux2024}
{Cadieux} C.,  et~al., 2024, \mn@doi [\apjl] {10.3847/2041-8213/ad5afa}, \href {https://ui.adsabs.harvard.edu/abs/2024ApJ...970L...2C} {970, L2}

\bibitem[\protect\citeauthoryear{Catling, Zahnle  \& McKay}{Catling et~al.}{2001}]{catling2001biogenic}
Catling D.~C.,  Zahnle K.~J.,   McKay C.~P.,  2001, Science, 293, 839

\bibitem[\protect\citeauthoryear{Catling et~al.,}{Catling et~al.}{2018}]{catlingdavid2018exoplanet}
Catling D.~C.,  et~al., 2018, Astrobiology, 18, 709

\bibitem[\protect\citeauthoryear{Chan et~al.,}{Chan et~al.}{2019}]{chan2019deciphering}
Chan M.~A.,  et~al., 2019, Astrobiology, 19, 1075

\bibitem[\protect\citeauthoryear{Charnay, Wolf, Marty  \& Forget}{Charnay et~al.}{2020}]{charnay2020faint}
Charnay B.,  Wolf E.~T.,  Marty B.,   Forget F.,  2020, Space Science Reviews, 216, 1

\bibitem[\protect\citeauthoryear{Chopra \& Lineweaver}{Chopra \& Lineweaver}{2016}]{chopra2016case}
Chopra A.,  Lineweaver C.~H.,  2016, Astrobiology, 16, 7

\bibitem[\protect\citeauthoryear{Chou et~al.,}{Chou et~al.}{2024}]{chou2024chapter}
Chou L.,  et~al., 2024, Astrobiology, 24, S

\bibitem[\protect\citeauthoryear{Christiansen}{Christiansen}{2022}]{christiansen2022five}
Christiansen J.~L.,  2022, Nature Astronomy, 6, 516

\bibitem[\protect\citeauthoryear{Cleaves, Hystad, Prabhu, Wong, Cody, Economon  \& Hazen}{Cleaves et~al.}{2023}]{cleaves2023robust}
Cleaves H.~J.,  Hystad G.,  Prabhu A.,  Wong M.~L.,  Cody G.~D.,  Economon S.,   Hazen R.~M.,  2023, Proceedings of the National Academy of Sciences, 120, e2307149120

\bibitem[\protect\citeauthoryear{Cleland}{Cleland}{2019a}]{cleland2019quest}
Cleland C.,  2019a, The Quest for a Universal Theory of Life: Searching for Life as we don't know it.
~ Vol. 11, Cambridge University Press

\bibitem[\protect\citeauthoryear{Cleland}{Cleland}{2019b}]{cleland2019moving}
Cleland C.~E.,  2019b, Astrobiology, 19, 722

\bibitem[\protect\citeauthoryear{Cleland \& Rimmer}{Cleland \& Rimmer}{2022}]{cleland2022ammonia}
Cleland C.~E.,  Rimmer P.~B.,  2022, Aerospace, 9, 752

\bibitem[\protect\citeauthoryear{Clements}{Clements}{2024}]{clements2024venus}
Clements D.~L.,  2024, arXiv preprint arXiv:2409.13438

\bibitem[\protect\citeauthoryear{Community}{Community}{2025}]{ldkb2025}
Community 2025, Life Detection Knowledge Base, \url {https://lifedetectionforum.com/ldkb}

\bibitem[\protect\citeauthoryear{Conan~Doyle}{Conan~Doyle}{1885}]{acdoyle_1885}
Conan~Doyle A.,  1885, Boys Own Paper

\bibitem[\protect\citeauthoryear{Fields, Gupta  \& Sandora}{Fields et~al.}{2023}]{fields2023information}
Fields B.,  Gupta S.,   Sandora M.,  2023, International Journal of Astrobiology, 22, 583

\bibitem[\protect\citeauthoryear{Foote, Sinhadc, Mathis  \& Walker}{Foote et~al.}{2023}]{foote2023false}
Foote S.,  Sinhadc P.,  Mathis C.,   Walker S.~I.,  2023, Astrobiology, 23, 1189

\bibitem[\protect\citeauthoryear{{Fressin} et~al.,}{{Fressin} et~al.}{2013}]{fressin2013}
{Fressin} F.,  et~al., 2013, \mn@doi [\apj] {10.1088/0004-637X/766/2/81}, \href {https://ui.adsabs.harvard.edu/abs/2013ApJ...766...81F} {766, 81}

\bibitem[\protect\citeauthoryear{Fujii et~al.,}{Fujii et~al.}{2018}]{fujii2018exoplanet}
Fujii Y.,  et~al., 2018, Astrobiology, 18, 739

\bibitem[\protect\citeauthoryear{Goldblatt}{Goldblatt}{2016}]{goldblatt2016inhabitance}
Goldblatt C.,  2016, arXiv preprint arXiv:1603.00950

\bibitem[\protect\citeauthoryear{Greaves et~al.,}{Greaves et~al.}{2021a}]{greaves2021reply}
Greaves J.~S.,  et~al., 2021a, Nature Astronomy, 5, 636

\bibitem[\protect\citeauthoryear{Greaves et~al.,}{Greaves et~al.}{2021b}]{greaves2021phosphine}
Greaves J.~S.,  et~al., 2021b, Nature Astronomy, 5, 655

\bibitem[\protect\citeauthoryear{Green, Hoehler, Neveu, Domagal-Goldman, Scalice  \& Voytek}{Green et~al.}{2021}]{green2021call}
Green J.,  Hoehler T.,  Neveu M.,  Domagal-Goldman S.,  Scalice D.,   Voytek M.,  2021, Nature, 598, 575

\bibitem[\protect\citeauthoryear{Greene, Bell, Ducrot, Dyrek, Lagage  \& Fortney}{Greene et~al.}{2023}]{greene2023thermal}
Greene T.~P.,  Bell T.~J.,  Ducrot E.,  Dyrek A.,  Lagage P.-O.,   Fortney J.~J.,  2023, Nature, 618, 39

\bibitem[\protect\citeauthoryear{{H{\"a}nni}, {Altwegg}, {Combi}, {Fuselier}, {De Keyser}, {Ligterink}, {Rubin}  \& {Wampfler}}{{H{\"a}nni} et~al.}{2024}]{nora2024}
{H{\"a}nni} N.,  {Altwegg} K.,  {Combi} M.,  {Fuselier} S.~A.,  {De Keyser} J.,  {Ligterink} N. F.~W.,  {Rubin} M.,   {Wampfler} S.~F.,  2024, \mn@doi [\apj] {10.3847/1538-4357/ad8565}, \href {https://ui.adsabs.harvard.edu/abs/2024ApJ...976...74H} {976, 74}

\bibitem[\protect\citeauthoryear{Harding \& Margulis}{Harding \& Margulis}{2010}]{harding2010water}
Harding S.,  Margulis L.,  2010, Gaia in turmoil: climate change, biodepletion, and Earth ethics in an age of crisis, pp 41--60

\bibitem[\protect\citeauthoryear{Heckert, Filliben, Croarkin, Hembree, Guthrie, Tobias  \& Prinz}{Heckert et~al.}{2002}]{heckert2002handbook}
Heckert N.~A.,  Filliben J.~J.,  Croarkin C.~M.,  Hembree B.,  Guthrie W.~F.,  Tobias P.,   Prinz J.,  2002, Handbook 151: Nist/sematech e-handbook of statistical methods.
NIST US Departement of Commerce, \mn@doi{https://doi.org/10.18434/M32189}

\bibitem[\protect\citeauthoryear{Hitchcock \& Lovelock}{Hitchcock \& Lovelock}{1967}]{hitchcock1967life}
Hitchcock D.~R.,  Lovelock J.~E.,  1967, Icarus, 7, 149

\bibitem[\protect\citeauthoryear{Kaltenegger, Lin  \& Madden}{Kaltenegger et~al.}{2020}]{kaltenegger2020high}
Kaltenegger L.,  Lin Z.,   Madden J.,  2020, The Astrophysical Journal Letters, 892, L17

\bibitem[\protect\citeauthoryear{Kasting}{Kasting}{2014}]{kasting2014}
Kasting J.~F.,  2014, in , Earth's Early Atmosphere and Surface Environment.
Geological Society of America

\bibitem[\protect\citeauthoryear{Kinney \& Kempes}{Kinney \& Kempes}{2022}]{kinney2022epistemology}
Kinney D.,  Kempes C.,  2022, Biology \& Philosophy, 37, 22

\bibitem[\protect\citeauthoryear{Kipping}{Kipping}{2025}]{kipping2025strong}
Kipping D.,  2025, arXiv preprint arXiv:2504.05993

\bibitem[\protect\citeauthoryear{Kopparapu, Ramirez, SchottelKotte, Kasting, Domagal-Goldman  \& Eymet}{Kopparapu et~al.}{2014}]{kopparapu2014habitable}
Kopparapu R.~K.,  Ramirez R.~M.,  SchottelKotte J.,  Kasting J.~F.,  Domagal-Goldman S.,   Eymet V.,  2014, The Astrophysical Journal Letters, 787, L29

\bibitem[\protect\citeauthoryear{Krissansen-Totton, Olson  \& Catling}{Krissansen-Totton et~al.}{2018}]{krissansen2018disequilibrium}
Krissansen-Totton J.,  Olson S.,   Catling D.~C.,  2018, Science advances, 4, eaao5747

\bibitem[\protect\citeauthoryear{Krissansen-Totton, Fortney, Nimmo  \& Wogan}{Krissansen-Totton et~al.}{2021}]{krissansen2021oxygen}
Krissansen-Totton J.,  Fortney J.~J.,  Nimmo F.,   Wogan N.,  2021, AGU Advances, 2, e2020AV000294

\bibitem[\protect\citeauthoryear{Lenton, Daines, Dyke, Nicholson, Wilkinson  \& Williams}{Lenton et~al.}{2018}]{lenton2018selection}
Lenton T.~M.,  Daines S.~J.,  Dyke J.~G.,  Nicholson A.~E.,  Wilkinson D.~M.,   Williams H.~T.,  2018, Trends in Ecology \& Evolution, 33, 633

\bibitem[\protect\citeauthoryear{Lingam, Haqq-Misra, Wright, Huston, Frank  \& Kopparapu}{Lingam et~al.}{2023}]{lingam2023technosignatures}
Lingam M.,  Haqq-Misra J.,  Wright J.~T.,  Huston M.~J.,  Frank A.,   Kopparapu R.,  2023, The Astrophysical Journal, 943, 27

\bibitem[\protect\citeauthoryear{Lovelock \& Margulis}{Lovelock \& Margulis}{1974}]{lovelock1974atmospheric}
Lovelock J.~E.,  Margulis L.,  1974, Tellus, 26, 2

\bibitem[\protect\citeauthoryear{Lowell}{Lowell}{1909}]{lowell1909mars}
Lowell P.,  1909, Science, 30, 338

\bibitem[\protect\citeauthoryear{{Madhusudhan}, {Constantinou}, {Holmberg}, {Sarkar}, {Piette}  \& {Moses}}{{Madhusudhan} et~al.}{2025}]{madhu2025}
{Madhusudhan} N.,  {Constantinou} S.,  {Holmberg} M.,  {Sarkar} S.,  {Piette} A. A.~A.,   {Moses} J.~I.,  2025, \mn@doi [\apjl] {10.3847/2041-8213/adc1c8}, \href {https://ui.adsabs.harvard.edu/abs/2025ApJ...983L..40M} {983, L40}

\bibitem[\protect\citeauthoryear{Matassa, Boon  \& Verstraete}{Matassa et~al.}{2015}]{matassa2015resource}
Matassa S.,  Boon N.,   Verstraete W.,  2015, Water research, 68, 467

\bibitem[\protect\citeauthoryear{McKay et~al.,}{McKay et~al.}{1996}]{mckay1996search}
McKay D.~S.,  et~al., 1996, Science, 273, 924

\bibitem[\protect\citeauthoryear{Meadows et~al.,}{Meadows et~al.}{2018}]{meadows2018exoplanet}
Meadows V.~S.,  et~al., 2018, Astrobiology, 18, 630

\bibitem[\protect\citeauthoryear{Meadows et~al.,}{Meadows et~al.}{2022}]{meadows2022community}
Meadows V.,  et~al., 2022, arXiv preprint arXiv:2210.14293

\bibitem[\protect\citeauthoryear{Meehl \& Rosen}{Meehl \& Rosen}{1955}]{meehl1955antecedent}
Meehl P.~E.,  Rosen A.,  1955, Psychological bulletin, 52, 194

\bibitem[\protect\citeauthoryear{Neveu, Hays, Voytek, New  \& Schulte}{Neveu et~al.}{2018}]{neveu2018ladder}
Neveu M.,  Hays L.~E.,  Voytek M.~A.,  New M.~H.,   Schulte M.~D.,  2018, Astrobiology, 18, 1375

\bibitem[\protect\citeauthoryear{Nicholson \& Mayne}{Nicholson \& Mayne}{2023}]{nicholson2023biotic}
Nicholson A.,  Mayne N.,  2023, Monthly Notices of the Royal Astronomical Society, 521, 5139

\bibitem[\protect\citeauthoryear{Nicholson, Daines, Mayne, Eager-Nash, Lenton  \& Kohary}{Nicholson et~al.}{2022}]{nicholson2022predicting}
Nicholson A.,  Daines S.,  Mayne N.,  Eager-Nash J.,  Lenton T.,   Kohary K.,  2022, Monthly Notices of the Royal Astronomical Society, 517, 222

\bibitem[\protect\citeauthoryear{Nisbet \& Sleep}{Nisbet \& Sleep}{2001}]{nisbet2001}
Nisbet G.~E.,  Sleep N.,  2001, Nature, 409, 1083

\bibitem[\protect\citeauthoryear{Popper}{Popper}{2005}]{popper2005logic}
Popper K.,  2005, The logic of scientific discovery.
Routledge

\bibitem[\protect\citeauthoryear{Ramirez}{Ramirez}{2018}]{ramirez2018more}
Ramirez R.~M.,  2018, Geosciences, 8, 280

\bibitem[\protect\citeauthoryear{{Reed}, {Shearer}, {McGlynn}, {Wing}, {Tolbert}  \& {Browne}}{{Reed} et~al.}{2024}]{reed2024}
{Reed} N.~W.,  {Shearer} R.~L.,  {McGlynn} S.~E.,  {Wing} B.~A.,  {Tolbert} M.~A.,   {Browne} E.~C.,  2024, \mn@doi [\apjl] {10.3847/2041-8213/ad74da}, \href {https://ui.adsabs.harvard.edu/abs/2024ApJ...973L..38R} {973, L38}

\bibitem[\protect\citeauthoryear{Sagan \& Mullen}{Sagan \& Mullen}{1972}]{sagan1972earth}
Sagan C.,  Mullen G.,  1972, Science, 177, 52

\bibitem[\protect\citeauthoryear{Sagan, Thompson, Carlson, Gurnett  \& Hord}{Sagan et~al.}{1993}]{sagan1993search}
Sagan C.,  Thompson W.~R.,  Carlson R.,  Gurnett D.,   Hord C.,  1993, Nature, 365, 715

\bibitem[\protect\citeauthoryear{Sandora \& Silk}{Sandora \& Silk}{2020}]{sandora2020biosignature}
Sandora M.,  Silk J.,  2020, Monthly Notices of the Royal Astronomical Society, 495, 1000

\bibitem[\protect\citeauthoryear{Schwartzman \& Volk}{Schwartzman \& Volk}{1991}]{schwartzman1991geophysiology}
Schwartzman D.~W.,  Volk T.,  1991, in Bioastronomy: The Search for Extraterrestrial Life—The Exploration Broadens. pp 155--162

\bibitem[\protect\citeauthoryear{Schwieterman et~al.,}{Schwieterman et~al.}{2018}]{schwieterman2018exoplanet}
Schwieterman E.~W.,  et~al., 2018, Astrobiology, 18, 663

\bibitem[\protect\citeauthoryear{Schwieterman, Reinhard, Olson, Ozaki, Harman, Hong  \& Lyons}{Schwieterman et~al.}{2019}]{schwieterman2019rethinking}
Schwieterman E.~W.,  Reinhard C.~T.,  Olson S.~L.,  Ozaki K.,  Harman C.~E.,  Hong P.~K.,   Lyons T.~W.,  2019, The Astrophysical Journal, 874, 9

\bibitem[\protect\citeauthoryear{Seager, Schrenk  \& Bains}{Seager et~al.}{2012}]{seager2012astrophysical}
Seager S.,  Schrenk M.,   Bains W.,  2012, Astrobiology, 12, 61

\bibitem[\protect\citeauthoryear{{Seager} et~al.,}{{Seager} et~al.}{2021}]{seager2021}
{Seager} S.,  et~al., 2021, \mn@doi [arXiv e-prints] {10.48550/arXiv.2112.05153}, \href {https://ui.adsabs.harvard.edu/abs/2021arXiv211205153S} {p. arXiv:2112.05153}

\bibitem[\protect\citeauthoryear{Sinton}{Sinton}{1957}]{sinton1957spectroscopic}
Sinton W.~M.,  1957, Astrophysical Journal, vol. 126, p. 231, 126, 231

\bibitem[\protect\citeauthoryear{Smith \& Mathis}{Smith \& Mathis}{2023}]{smith2023life}
Smith H.~B.,  Mathis C.,  2023, BioEssays, 45, 2300050

\bibitem[\protect\citeauthoryear{Sol{\'e} \& Munteanu}{Sol{\'e} \& Munteanu}{2004}]{sole2004large}
Sol{\'e} R.~V.,  Munteanu A.,  2004, Europhysics Letters, 68, 170

\bibitem[\protect\citeauthoryear{Spiegel \& Turner}{Spiegel \& Turner}{2012}]{spiegel2012bayesian}
Spiegel D.~S.,  Turner E.~L.,  2012, Proceedings of the National Academy of Sciences, 109, 395

\bibitem[\protect\citeauthoryear{Styczinski, Cooper, Glaser, Lehmer, Mierzejewski  \& Tarnas}{Styczinski et~al.}{2024}]{styczinski2024chapter}
Styczinski M.,  Cooper Z.,  Glaser D.,  Lehmer O.,  Mierzejewski V.,   Tarnas J.,  2024, Astrobiology, 24, S

\bibitem[\protect\citeauthoryear{Tarter}{Tarter}{2001}]{tarter2001search}
Tarter J.,  2001, Annual Review of Astronomy and Astrophysics, 39, 511

\bibitem[\protect\citeauthoryear{{Taylor}}{{Taylor}}{2025}]{taylor2025}
{Taylor} J.,  2025, \mn@doi [arXiv e-prints] {10.48550/arXiv.2504.15916}, \href {https://ui.adsabs.harvard.edu/abs/2025arXiv250415916T} {p. arXiv:2504.15916}

\bibitem[\protect\citeauthoryear{{Turbet}, {Ehrenreich}, {Lovis}, {Bolmont}  \& {Fauchez}}{{Turbet} et~al.}{2019}]{turbet2019}
{Turbet} M.,  {Ehrenreich} D.,  {Lovis} C.,  {Bolmont} E.,   {Fauchez} T.,  2019, \mn@doi [\aap] {10.1051/0004-6361/201935585}, \href {https://ui.adsabs.harvard.edu/abs/2019A&A...628A..12T} {628, A12}

\bibitem[\protect\citeauthoryear{{Turbet} et~al.,}{{Turbet} et~al.}{2023}]{turbet2023}
{Turbet} M.,  et~al., 2023, \mn@doi [\aap] {10.1051/0004-6361/202347539}, \href {https://ui.adsabs.harvard.edu/abs/2023A&A...679A.126T} {679, A126}

\bibitem[\protect\citeauthoryear{Vickers, Cowie, Dick, Gillen, Jeancolas, Rothschild  \& McMahon}{Vickers et~al.}{2023}]{vickers2023confidence}
Vickers P.,  Cowie C.,  Dick S.~J.,  Gillen C.,  Jeancolas C.,  Rothschild L.~J.,   McMahon S.,  2023, Astrobiology, 23, 1202

\bibitem[\protect\citeauthoryear{Villanueva et~al.,}{Villanueva et~al.}{2021}]{villanueva2021no}
Villanueva G.~L.,  et~al., 2021, Nature Astronomy, 5, 631

\bibitem[\protect\citeauthoryear{Walker et~al.,}{Walker et~al.}{2018}]{walker2018exoplanet}
Walker S.~I.,  et~al., 2018, Astrobiology, 18, 779

\bibitem[\protect\citeauthoryear{Watson \& Lovelock}{Watson \& Lovelock}{1983}]{watson1983biological}
Watson A.~J.,  Lovelock J.~E.,  1983, Tellus B: Chemical and Physical Meteorology, 35, 284

\bibitem[\protect\citeauthoryear{Wogan \& Catling}{Wogan \& Catling}{2020}]{wogan2020chemical}
Wogan N.~F.,  Catling D.~C.,  2020, The Astrophysical Journal, 892, 127

\bibitem[\protect\citeauthoryear{{Wogan}, {Batalha}, {Zahnle}, {Krissansen-Totton}, {Tsai}  \& {Hu}}{{Wogan} et~al.}{2024}]{wogan2024}
{Wogan} N.~F.,  {Batalha} N.~E.,  {Zahnle} K.~J.,  {Krissansen-Totton} J.,  {Tsai} S.-M.,   {Hu} R.,  2024, \mn@doi [\apjl] {10.3847/2041-8213/ad2616}, \href {https://ui.adsabs.harvard.edu/abs/2024ApJ...963L...7W} {963, L7}

\bibitem[\protect\citeauthoryear{Wong, Prabhu, Williams, Morrison  \& Hazen}{Wong et~al.}{2023}]{wong2023toward}
Wong M.~L.,  Prabhu A.,  Williams J.,  Morrison S.~M.,   Hazen R.~M.,  2023, Journal of Geophysical Research: Planets, 128, e2022JE007658

\bibitem[\protect\citeauthoryear{Wong, Duckett, Hernandez, Rajaei  \& Smith}{Wong et~al.}{2024}]{wong2024process}
Wong M.~L.,  Duckett M.,  Hernandez E.~S.,  Rajaei V.,   Smith K.~J.,  2024, Perspectives of Earth and Space Scientists, 5, e2023CN000223

\bibitem[\protect\citeauthoryear{Wright, Haqq-Misra, Frank, Kopparapu, Lingam  \& Sheikh}{Wright et~al.}{2022}]{wright2022case}
Wright J.~T.,  Haqq-Misra J.,  Frank A.,  Kopparapu R.,  Lingam M.,   Sheikh S.~Z.,  2022, The Astrophysical Journal Letters, 927, L30

\makeatother
\end{thebibliography}








\bsp	
\label{lastpage}
\end{document}